\documentclass[showpacs,preprintnumbers,superscriptaddress,amsmath,amssymb,prl]{revtex4}

\usepackage{graphicx} 
\usepackage{dcolumn} 
\usepackage{bm} 
\usepackage{amsmath}
\usepackage{indentfirst} 
\usepackage{url}

\newcommand{\be}{\begin{equation}} 
\newcommand{\ee}{\end{equation}}
\newcommand{\ba}{\begin{eqnarray}} 
\newcommand{\ea}{\end{eqnarray}}

\begin{document}

\preprint{APS preprint}

\title{Electro-Magnetic Earthquake Bursts \\and Critical Rupture of Peroxy
Bond Networks in Rocks}

\author{F.~Freund} 
\affiliation{Department of Physics, San Jose State
University San Jose, CA 95192-0106 and NASA Ames Research Center, Mail
Stop 242-4, 515 N. Whisman Road, CA 94035-1000 and SETI Institute, 2035
Landing Drive,	Mountain View, CA 94043, USA (e-mail:
ffreund@mail.arc.nasa.gov)}

\author{D. Sornette} 
\affiliation{Chair of Entrepreneurial Risks, D-MTEC.
 ETH Zurich (Swiss Federal Institute of Technology Zurich), CH-8032 Zurich 
Switzerland (email: dsornette@ethz.ch)}
\affiliation{Laboratoire de Physique de la Mati\`ere Condens\'ee, CNRS UMR6622 
and Universit\'e des Sciences, Parc Valrose
06108 Nice Cedex 2, France}

\date{\today}

\begin{abstract} 
We propose a mechanism for the low frequency electromagnetic emissions
and other electromagnetic and electric phenomena
which have been associated with earthquakes.
The mechanism combines the critical earthquake concept and
the concept of crust acting as a charging electric battery under
increasing stress. The electric charges are released by activation of
dormant charge carriers in the oxygen anion sublattice, called peroxy
bonds or positive hole pairs (PHP), where a PHP represents an
$O_3X/^{OO}\backslash YO_3$ with $X,Y = Si^{4+}, Al^{3+}...$, i.e.
$O^-$ in a matrix of $O^{2-}$ of silicates. We propose that PHP are
activated by plastic deformations during the slow cooperative
build-up of stress and the increasingly correlated damage culminating in
a large ``critical'' earthquake. Recent laboratory experiments indeed
show that stressed rocks form electric batteries which can release their
charge when a conducting path closes the
equivalent electric circuit. We conjecture that the intermittent and
erratic occurrences of EM signals are a consequence
of the progressive build-up of the battery charges 
in the Earth crust and their erratic release when
crack networks are percolating throughout the stressed rock
volumes, providing a conductive pathway for the battery currents to
discharge. EM signals are thus expected close to the rupture, either
slightly before or after, that is, when percolation
is most favored. The proposed mechanism should be relevant for the
broader understanding of fractoemissions.
\end{abstract}

%\pacs{91.30.Px ; 89.75.Da; 05.40.-a}

\maketitle

\section{Introduction}

There are many reports of electric and electro-magnetic radiations associated with earthquakes,
The observations include (NASDA-CON-960003, 1997; Johnston, 1997)
low-frequency electromagnetic emissions (Fujinawa et al., 2001;
Fujinawa and Takahashi, 1990; Gershenzon and Bambakidis, 2001; Gokhberg
et al., 1982; Molchanov and Hayakawa, 1998a,b; Nitsan, 1977; Vershinin et
al., 1999; Yoshida et al., 1994; Yoshino and Tomizawa, 1989; Taylor and Purucker, 2002), local
magnetic field anomalies (Fujinawa and Takahashi, 1990; Gershenzon and
Bambakidis, 2001; Ivanov et al., 1976; Kopytenko et al., 1993; Ma et
al., 2003; Yen et al., 2004; Zlotnicki and Cornet, 1986), increases in
radio-frequency noise (Bianchi and al., 1984; Hayakawa, 1989; Martelli
and Smith, 1989; Pulinets et al., 1994), earthquake lights (Derr,
1973; King, 1983; St-Laurent, 2000; Tsukuda, 1997; Yasui, 1973), seismic
electric signals (Varotsos, 2005; Debate on VAN, 1996) and so on.
The ability of the Earth's crust in areas of active deformation to transform
mechanical energy into electromagnetic radiations may be due to one or several
mechanisms. The mechanisms which have been proposed in the literature include
piezoelectric effects, streaming potentials, Lenard splashing, plasma
generation, electron emission, separating electrification,
pseudo-piezoelectric effects, crack surface friction and so on. 
Here, we examine in some details the mechanism based on
the release of electric charges by activation of
dormant charge carriers in the oxygen anion sublattice, called peroxy
bonds or positive hole pairs (PHP). 

We first present a short review of the hypothesis proposed by
the first author and his co-authors that electric charges 
are activated in rocks as stress
increases, making the crust an electric battery which can discharge.
The discharge currents may
lead to electromagnetic (EM) emissions associated with the
closing of the electric circuit by the percolation of conducting paths most
probably occurring around the time of (before, during and after)
large earthquakes. We then briefly review the critical earthquake concept
proposed some time ago (Sornette and Sornette, 1990; Sornette and Sammis, 1995;
Saleur et al., 1996, Bowman et al., 1998; Huang et al., 1998; Sammis and
Sornette, 2002), which views large earthquakes as the
culmination of a growing collective organization of damage similar to a
critical phase transition (see the recent review (Sornette, 2005) for its
description in the context of material rupture). We then combine the two
concepts to propose an explanation for why powerful low frequency EM
emissions would become observable under certain conditions but remain
unobservable in other cases. Eftaxias et al. have previously found that EM
signals reveal characteristic signs of an approaching critical point
(Eftaxias et al., 2002; 2004; Kapiris et al., 2004a,b; Contoyiannis et al.,
2005). Here, we continue along this line of thought by proposing a
specific mechanism for electric charge generation at the basis of EM
emissions, which is combined with the theory of a critical organization
of the crust before large earthquakes. This allows us to suggest some
observational consequences for EM precursors of earthquakes.

\section{Incipient damage, peroxy bonds and positives holes}

Recent work has brought forth evidence of electric signatures
associated with mechanical disturbances 
(Freund, 2000; Freund et al., 2006a,b). These experiments
have led to the following understanding that we briefly summarize.

The first step is to
recognize that electronic charge carriers exist in igneous and
high-grade metamorphic rocks in a dormant, electrically inactive form. 
These charge carriers are defect electrons in the oxygen anion
sublattice, also known as
positive holes, and their dormant precursors
are positive hole pairs (PHP) (see Figure \ref{fig1}). Among the mechanisms that activate the
PHPs and generate the highly mobile positive holes are processes that
are expected to occur on a large scale in rock volumes experiencing an
increasing stress due to tectonic forces. The two main processes are
plastic flow and microfracturing, both that are thought 
to occur ubiquitously in the earth crust. During plastic 
deformation that underlies plastic flow,
dislocations, which sweep through the crystal structures of the constituent
minerals, pass through PHPs and cause them instently to release
positive hole charge carriers. The stress field associated with
microfractures can cause additional activation of PHPs within the surrounding
microvolume. In a rock experiencing massive compressive stress such as
in thrust setting characteristic of subduction zones or in shear zones
under global tectonic forces, these processes are predicted to lead to a
large-scale activation of positive hole charge carriers that would flow
out of the ``source volume''. Both, plastic deformation and
microfracturing of rocks are expected to mark the on-going preparation
of rock deformation towards global failure, i.e. prior to initiation of
an earthquake.

Let us now step back for a while to justify the key role that PHP, we
believe, play in the signature of large damage processes. Martens et
al. (1976) reported early that hydroxyl pairs in a model mineral, $MgO$,
converted to molecular $H_2$ while oxidizing two regular lattice oxygen
anions from the 2- to the 1- state, hence to peroxy:  
\be
OH\bar{ } + OH\bar{ } \Leftrightarrow H_2 + O_2^{2}\bar{ }~.
\ee
This redox conversion was confirmed spectroscopically (Freund and Wengeler,
1982) and through electrical conductivity and dielectric polarization
studies (Freund et al., 1993). The same redox conversion was found in
olivine which, though nominally anhydrous, contains traces of ``water''
in form of hydroxyl, $O_3XÐOH$, where $X=Si^{4+}, Al^{3+}$, and so on (Freund
and Oberheuser, 1986). It has long been recognized that all nominally
anhydrous minerals (meaning nearly all minerals in common igneous rocks)
contain traces of dissolved ``water'' (Bell and Rossman, 1992; Martin
and Wyss, 1975; Rossman, 1996; Smyth et al., 1991 and revue of Sornette, 1999).  
What had not been
recognized is that $XÐOH$ pairs in these minerals appear to undergo the
same redox reaction as $OH\bar{ }$ pairs in $MgO$, giving rise to peroxy
links:  
\be
O_3XÐOH...HOÐXO_3 \Leftrightarrow H_2 + O_3X/^{OO}\backslash XO_3~.
\ee
This reaction occurs below $400-500^{\circ}$C. The geoscience
community at large has not yet taken notice of this redox conversion. 
One of the prime reason seems to be the assumption that oxygen is never
in any other oxidation state but $2$-.  Accepting the possibility of
oxygen in the $1$- oxidation state in peroxy links opens a road towards
understanding earthquake-related electrical phenomena.
Figure \ref{fig1} presents a schematic representation of these
reactions corresponding to the conversion of hydroxyl pairs in MgO to
peroxy (PHP) plus $H_2$ and the ``tearing apart'' of the PHP to generate
p-holes. As argued below, we propose that analogous reactions occur
for Si-OO-Si in silicates.

Peroxy and positive holes (positive electronic charge carriers) are defect
electrons or ``holes.''  When occurring in the $O^{2}\bar{ }$ sublattice of
oxides or silicate minerals, they are called ``positive holes'' 
or p-holes for short and designated $h\dot{ }$. Chemically, 
a positive hole is an $O\bar{ }$. While an
$O\bar{ }$ is a radical, two $O\bar{ }$ can stabilize by pairing and tying an
$O^ÐÐO^Ð$ bond. An $O^ÐÐO^Ð$ bond is a peroxy link,
$O_3X/^{OO}\backslash YO_3$ with $X,Y = Si^{4+}, Al^{3+}$ and so on. As we
have said, in physical terms, a peroxy link is a positive hole pair (PHP).

Magmatic and high-grade metamorphic rocks consist primarily or entirely of
nominally anhydrous minerals that crystallized or recrystallized in
$H_2O$-laden environments. Invariably, such minerals incorporate small
amounts of $H_2O$.  The dissolved water occurs in form of hydroxyls,
e.g. as $OH\bar{ }$ or $XÐOH$ (Bell and Rossman, 1992; Martin and Wyss, 1975;
Rossman, 1996; Smyth et al., 1991): 
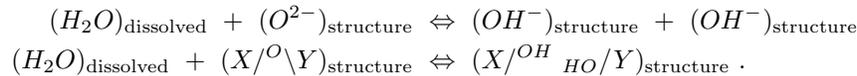
\ba 
(H_2O)_{\rm dissolved}~  + 
~(O^{2-})_{\rm structure} & \Leftrightarrow & (OH^-)_{\rm structure}~ +
~ (OH^-)_{\rm structure} \nonumber \\ 
(H_2O)_{\rm dissolved}~  + ~(X/^O\backslash Y)_{\rm structure} 
& \Leftrightarrow & (X/^{OH}~
_{HO}/Y)_{\rm structure} 	~.  
 \nonumber 
 \ea 
A major fraction of
these hydroxyls is believed to undergo this
particular redox conversion by which $OH^-$ or $X-OH$
pairs rearrange so as to reduce two $H^+$ to $H_2$ and oxidize two
$O^{2-}$ to $O^-$ which tie a peroxy link: 
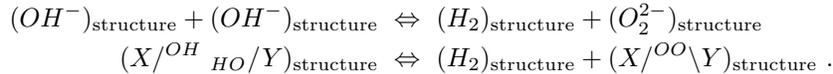
\ba 
(OH^-)_{\rm structure} +
(OH^-)_{\rm structure}  &\Leftrightarrow& (H_2)_{\rm structure}  + 
(O_2^{2-})_{\rm structure}	\nonumber \\ 
(X/^{OH}~ _{HO}/Y)_{\rm
structure}  & \Leftrightarrow &   (H_2)_{\rm structure} + 
(X/^{OO}\backslash Y)_{\rm structure}~.   \nonumber 
\ea 
Even minerals
that come from highly reducing crustal and upper mantle environments and
contain reduced transition metal cations can acquire peroxy.  This may
appear paradoxical but peroxy and reduced transition metals are not
mutually exclusive. This is because the redox conversion takes place
during cooling, at relatively low temperatures, below
$400-500^{\circ}$C. At these low temperatures, all major solid state
processes are frozen and non-equilibrium conditions prevail (Kathrein et
al., 1984; Sornette, 1999).

Basic information about PHP's and positive hole charge carriers, $h\dot{ }$,
was obtained by studying $MgO$, theoretically (King and Freund, 1984)
and experimentally by dc conductivity, dielectric polarization
(Freund et al., 1993), thermal
expansion, refractive index measurements etc. (Freund et al., 1994) and
fracture experiments (Dickinson et al., 1986).  Information about PHP's
in upper mantle olivine was derived from dielectric polarization and
fracture mass spectroscopy experiments (Freund and Ho, 1996). Peroxy
links in fused $SiO_2$ were studied theoretically by Edwards and Fowler
(1982) and Ricci et al. (2001), 
and experimentally by electron spin resonance spectroscopy
(Friebele et al., 1979) and dielectric polarization (Freund et al.,
1993). Fracture experiments with $MgO$ crystals suggested that the
stress fields and the acoustic waves emitted during fracture activate
$h\dot{ }$ charge carriers (Dickinson et al., 1986; Freund et al., 2002). 
Recent experiments using
slow velocity impacts have shown the validity of the concept that PHPs
are activated under the effect of time-dependent stresses 
(Freund, 2002). The experiments used steel ball bearings
impacting at $100$ m/sec on quartz-free gabbro ($\approx 80\%$
plagioclase, $\approx 20\%$ pyroxene) and low-quartz diorite cut to
cylindrical cores.  Sensors using magnetic pick-up coils, photodiodes,
capacitive sensors and contact electrodes clearly demonstrated the
appearance of positive carriers that were created by the impact. The conditions
under impact resemble those experienced microscopically by the mineral
grains in a large rock volume that is slowly compressed with
intermittent damage. Though the overall stress change may be small, the
strain will tend to discharge ``explosively'' on the
microscopic scale by bursts of dislocation
movements and microfractures that open and close on short time scales.
These bursts, though numerous, occur distributed over time and space.
Another recent experiment by Freund et al. (2006b)
shows that, when stress is applied to one end of a block of igneous
rocks, two currents flow out of the stressed rock volume. One current is
carried by electrons and it flows out through a Cu electrode directly
attached to the stressed rock volume. The other current is carried by
p-holes, i.e. defect electrons on the oxygen anion sublattice, and it
flows out through at least 1 m of unstressed rock. The two currents are
coupled via their respective electric fields and fluctuate. Evidence
of these two types of currents include the positive sign of the charges accumulating
on the rock surface which can be directly measured. In sum, the 
stressed rock volume acts as a genuine electric battery, which can discharge only
when the two poles are connected via a conductor, i.e., when
the electric circuit is closed on itself.

Dry rocks are usually good insulators but become p-type semiconductors
as a consequence of the break-up of the peroxy bonds. The p-hole charge
carriers have not been recognized earlier probably due to the fact that
a $h\dot{ }$ is (chemically speaking) an $O\bar{ }$ which is highly oxidizing and
reacts with any reduced gas and is thus annihilated. All the
studies on the electrical conductivity of minerals and rocks noticed the
unusually high conductivity in the $400-600^{\circ}$C window but assumed
it away as a result of surface contamination; indeed, in all these studies,
the anomaly was eliminated by annihilating the $h\dot{ }$ charge carriers by
``equilibrating'' their minerals and rocks for long time in reduced gas
until the conductivity returned to ``normal''!

Applying the
insight gained from these laboratory experiments
(Freund, 2002; Freund et al., 2006b) to the field, where
large volume of rocks are subjected to ever increasing stress, has led us
to suggest the existence of transient, fluctuating currents of considerable magnitude
that could build up in the Earth's crust prior to major earthquakes.
An important problem however is to understand how large scale currents
can develop and under what circumstances. For this up-scaling purpose, we propose
the critical earthquake model, that we now briefly review before
combining it with the ``battery'' hypothesis.

\section{The critical-earthquake model}

As often in earthquake physics, we face the challenge of up-scaling,
i.e., the transition from the laboratory scale to the scale of the
Earth crust. In order to explain EM signals which appear at large scales,
it seems necessary to
identify {\it global} signatures of the cooperative damage occurring at
small scales. This leads us to proposing the relevance of the critical-earthquake model.
We emphasize that a local analysis is inherently
incompatible with the critical earthquake hypothesis. The problem is similar to the
prediction of incipient failure in materials in the laboratory or in
engineering structures (Anifrani et al., 1995; Johansen and Sornette,
2000): only by integrating the information on many damaged elements
can one develop efficient diagnostic for failure (Andersen and Sornette,
2005; Sornette and Andersen, 2006), in contrast with the detailed
description of individual cracking events which are like the trees
hiding the forest. Here, we describe a physical mechanism that converts
damage at the micro-scale into electric signals, such that the
cooperative and progressive damage occurring over a large spatial region
preceding the main shock should translate into large-scale observable
electro-magnetic signatures.

We start with a cartoon of the mechanical processes occurring
under stress, first on a small scale of individual grains, then on a
larger scale within the crust taking into account the variation of
essential parameters such as temperature, ductility and electrical
conductivity. Figure \ref{fig2} shows schematically the response of a cube of
rock, confined within a larger rock volume and subjected to increasing
stress, (a) if the sample is brittle, and (b) if the sample is ductile.

If a brittle sample is subjected to stress beyond its elastic limit, the
deformations become nonlinear. On an atomic scale, dislocations move and
new dislocations are generated. As the dislocations per unit volume
become ever more numerous, they begin to coalesce, leading to
microfractures. As the microfractures become ever more numerous, they
too coalesce leading to larger cracks. These larger cracks will
eventually lead to catastrophic failure.

In a ductile sample, the dislocations anneal as fast as they are
generated under stress. The result is plastic flow without
microfractures and, hence, without cracks and catastrophic failure. It
is important to note, however, that even a ductile material can become
brittle and develop fractures, if the stresses reach very high values.
The reason is that the brittle-ductile transition is controlled by two
time-dependent processes: by the rate at which dislocations are generated
and by the rate at which dislocations anneal. If, at high stress rates,
more dislocations are generated per unit time and unit volume than can
be annealed, dislocations can pile up and coalesce even in a ductile
material, leading to microfractures and cracks. This is graphically
represented on the right hand side of Figure \ref{fig2}b.

In Figure \ref{fig3}, we apply this small-scale concept to infer
the qualitative mechanical behavior of favorable domains for the nucleation
and propagation of future earthquakes. We keep in mind the fact that the stress
field is probably highly heterogeneous (Shamir et al., 1990; 
Hickman et al., 2000). Figure \ref{fig3} represents the effective
rheology of a large sample converging to rupture. In the upper left
of Figure \ref{fig3}, we show a cross section through two crustal blocks which we
assume to have collided at time $t_0$ along the dashed line. The two blocks
comprise the cool, brittle upper crust and the increasingly hot, ductile
mid- to lower crust. We further assume that, due to tectonic forces, the
two blocks are being pushed against each other at a constant speed
(constant strain). Also shown within the blocks are two cross-hatched
zones: one where the rocks change from brittle to ductile and another one
where their conductivity changes from p-type to n-type. Indeed, 
we include one important electrical parameter: the fact
that rocks in the cool upper crust tend to be ever so slightly p-type
conducting but turn n-type conducting deeper into the crust when the
temperatures reach or exceed $500-600^{\circ}$C (Freund, 2003). 

In Figure \ref{fig3}a, we show that the strain increases linearly with time. In
Figure \ref{fig3}b we plot in a simplified way strain versus stress. In the
elastic range, the strain is linear with stress. In the inelastic range,
which is of interest here, the strain, i.e. deformation, increases
non-linearly, due to the generation and movement of dislocations.
Eventually, under the onslaught of continuing deformation at a constant
strain rate, stress increases even more rapidly and the system becomes
mechanically unstable. In Figure \ref{fig3}c we plot, again schematically, how
the viscosity changes with increasing depth and, hence, with increasing
temperature. The deeper we go into the crust, the more the temperature
increases. Hence, the rocks become ever more ductile. The brittle to
ductile transition is marked by a cross-hatched region. However, the
position of the brittle-to-ductile transition zone is not fixed. The
reason is, as explained in the context of Figure \ref{fig2}b, a ductile response
at low stress levels (slow deformation) may turn into a brittle response
when high levels of stress are applied (rapid deformation).

Finally, in Figure \ref{fig3}d, we combine the information derived from Figures
\ref{fig3}a-c with the representation of two crustal blocks colliding at a
constant speed. In Figure \ref{fig3}d, we plot the deformation volume versus time,
i.e. the rock volume that reaches a certain level of deformation as time
progresses. We can arbitrarily choose as the level of deformation the
on-set of microfracturing as depicted in Figure \ref{fig3}a. Because of the
non-linear increase of stress (Figure \ref{fig3}b), combined with the decrease in
viscosity (Figure \ref{fig3}c), we know that the deformation volume, which
satisfies this condition, will expand outward and downward with
increasing stress as indicated by the white lines in the two-block
model (Ashby and Jones, 1980; Frost and Ashby, 1982; Nechad et al., 2005a,b). 
The points $t_0$, $t_1$ and $t_2$ on the time axis of Figure \ref{fig3}d mark the
initial stress build-up. We assume that the deformation volume reaches
the transition zone from brittle to ductile at $t_3$ and the transition
zone from p-type to n-type behavior at $t_4$.

While we expect the brittle-to-ductile transition zone to move downward
in the crust when stresses become high and deformations become rapid,
the transition from p-type to n-type is controlled by temperature only
and independent of the stress level (Freund 2003). Thus, we can envision
situations where, with ever increasing stress levels, the transition to
a ductile behavior is pushed deeper and deeper into the crust. This
would allow the zone of brittle fracturing to extend downward,
eventually overlapping with the zone where the rocks become pervasively
n-type. At this point, we believe, a crucial electrical connection is
made which enables the electrons in the battery, i.e. in the stressed
rock volume, to massively flow downward and create a condition where the
p-hole currents follow suit by also flowing out massively.

Uncorrelated percolation (Stauffer and Aharony,
1992) provides a starting modeling point valid in the limit of very
large disorder (Roux et al., 1988). For realistic systems, long-range
correlations transported by the stress field around defects and cracks
make the problem much more subtle. Time dependence is expected to be a
crucial aspect in the process of correlation building in these
processes. As the damage increases, a new ``phase'' appears, where
micro-cracks begin to merge leading to screening and other cooperative
effects. This should lead to an overall acceleration of the seismicity
at large scales. However, by the very nature of the damage processes,
this acceleration is expected to be very intermittent and strongly
fluctuating from case to case. With respect to both 
mechanical and electric (and therefore electro-magnetic) precursors,
the big unknown is still the role of water in gouge-filled faults or
aquifers. In addition, given the complexity of the crust, large-scale
processes will most likely be influenced as well by factors such as the
types of rocks in the stressed volume. Finally, the main fracture is
formed causing global failure. This scenario
is very different from the nucleation model (Dieterich, 1992), which
envisions instead a local preparation or nucleation zone with a typical
size of meters to maybe a few kilometers at most. Within the nucleation
model, precursors are not expected to occur 
over very large spatial regions.

The idea, that great earthquakes could be ``critical'' events in a
technical sense of the term explained below, is gaining momentum in the
geophysical community, even if it is far from proven as its consequences
are still debated. The critical earthquake model
provides an original view of the organization of the crust prior to
great earthquakes that, as we shall see, allows one to explain several
paradoxical observations, which pose problem within the 
nucleation model (Dieterich, 1992).

The word ``critical'' describes a system at the boundary between order
and disorder, that is characterized by both extreme susceptibility to
external factors and long-range correlations between different parts of
the system (Sornette, 2004). Examples of such systems are 
magnets close to their Curie point, where the system progressively
orders under small external changes.  Critical behavior is fundamentally
a cooperative phenomenon, resulting from the repeated interactions
between ``microscopic'' elements which progressively ``phase up'' and
construct a ``macroscopic'' self-similar state.

This idea can be traced back to the critical branching model of
Vere-Jones (1977) and to the percolation model of rupture using the
real-space renormalization group approach (All\`egre et al., 1982). The
Russian school has also extensively developed this concept  (see
(Keilis-Borok, 1990) and references therein). Sornette and Sornette
(1990) and Sornette and Sammis (1995) identified specific measurable
signatures of the predicted critical behavior in terms of a
time-to-failure power-law acceleration of physical properties such as
electric precursors (Sornette and Sornette, 1990) or the Benioff strain,
previously interpreted differently with empirical mechanical-damage laws
(Sykes and Jaum\'e, 1990; Bufe and Varnes, 1993).

The critical-earthquake concept has been further strengthened by showing
that a strong heterogeneity of the elastic and/or
failure properties of rocks is necessary for a critical behavior of
rupture to occur (Andersen et al., 1997; Heimpel, 1997; Sornette and Andersen,
1998; Lei et al., 2000; 2004).
This was anticipated early by Mogi
(1969), who noticed that, for experiments on a variety of materials, the
larger the disorder, the stronger and more useful are the precursors to
rupture. In a heterogeneous material, the fracture behavior will be
determined by the block that is most brittle and fails in a catastrophic
way. The disorder may not need to be only frozen (pre-existing) 
as it may be generated during the deformation processes (Bouchaud and Mezard, 1994),
in particular in the transition from the generation/movement of dislocations to
coalescence/microfracturing. 
Numerical simulations have confirmed that, near the global
failure point, the cumulative elastic energy released during fracturing
of heterogeneous solids with long-range elastic interactions exhibit a
critical behavior with observable log-periodic corrections (Sahimi and
Arbati, 1996; Johansen and Sornette, 1998; Zhou and Sornette, 2002).
Recent experiments measuring acoustic emissions associated with the
rupture of fiber-glass composites (Garcimartin et al., 1997; Moura and
Yukalov, 2002; Yukalov et al., 2004, Nechad et al., 2005a,b), on kevlar
and carbon fiber composites (Johansen and Sornette, 2000) and on rocks
(Lei et al., 2003; Moura et al., 2005) have confirmed the critical
scenario. Sammis and Sornette (2002) have generalized the ``critical
point'' concept for large earthquakes in the framework of so-called
``finite-time singularities,'' and have proposed that the singular
behavior associated with accelerated seismic release results from a
positive feedback of the seismic activity on its release rate. The most
important mechanisms for such positive feedback include (i) stress load
in subcritical crack growth, (ii) geometrical effects in creep rupture,
(iii) weakening by damage, (iv) stress redistribution in a percolation
model of asperity failures and (v) fragmentation of the stress-shadow
(cast by the last large earthquake) by the increasing tectonic stress. The
critical-earthquake model has also been tested with positive results on
rockbursts occurring in deep South African mines (Ouillon and Sornette,
2000), thus offering an example with an intermediate range of scales
between the laboratory and the earth crust.

As predicted by the critical earthquake model, there are many reported
observations of increased intermediate magnitude seismicity before large
events  (Ellsworth et al., 1981; Jones, 1994; Keilis-Borok et al., 1988;
Knopoff et al., 1996; Lindh, 1990; Main, 1996; Mogi, 1969; Raleigh et
al., 1982; Sykes and Jaum\'e, 1990; Tocher, 1959).  Because these
precursory events occur over an area much greater than is predicted for
elasto-dynamic interactions, they are not considered to be classical
foreshocks (Jones and Molnar, 1979).  While the observed long-range
correlations in seismicity can not be explained by simple elasto-dynamic
interactions, they can be understood by analogy to the statistical
mechanics of a system approaching a critical point for which the
correlation length is only limited by the size of the system (Wilson,
1979). There are large fluctuations of the precursory
seismicity from event to event. Therefore, the statistical significance
of these precursors has not yet been established firmly (Huang et al.,
2000, Helmstetter and Sornette, 2003).
Another ingredient for large scale correlations of the precursory
signals is that the tectonic plates are being
moved or dragged along by large-scale mantle convections and,
hence, stresses act coherently on large sections of the crust. 

If the crust does behave such that larger earthquakes are ``critical''
events, stress correlation lengths should grow in the lead-up to large
events and drop sharply once these occur.  In the critical model, a
large or great earthquake dissipates a sufficient proportion of the
accumulated regional strain to destroy these long wavelength stress
correlations, which need to be re-established to prepare for the next
big one. According to this model, large earthquakes are not just
scaled-up version of small earthquakes but play a special role as
``critical points'' (Bowman et al., 1998; Brehm and Braile, 1999a,b; Jaume
and Sykes, 1999). However, this evolution in stress correlation lengths
is very difficult to observe directly. Recently, Mora and Place (2000)
have shown, using the lattice solid model to describe discontinuous
elasto-dynamic systems subjected to shear and compression presented
in (Place et al., 1999), that cumulative strain and correlation
lengths do exhibit a critical evolution in these model systems. Huang et
al. (1998) have shown that the concept of critical earthquake is
compatible with the concept of self-organized criticality (Bak and Tang,
1989; Sornette and Sornette, 1989) in a hierarchical structured crust
(Ouillon et al., 1996). This has been further elaborated in terms of the
concept of ``intermittent criticality'' (Bowman and Sammis, 2004; Sammis
et al., 2004). 

An alternative proposal, called the Stress Accumulation
model sees the accelerating seismicity sequences as resulting from the
tectonic loading of large fault structures through aseismic slip in the
elasto-plastic lower crust (Bowman and King, 2001a,b; King and Bowman, 2003; Levin et al.,
2006). According to this view, the upper
crust could be driven in significant part from below (lower crust and
mantle) with its associated plastic and plastic-ductile localization
controlling in part the organization of faulting (Regenauer-Lieb and
Yuen, 2003) and of seismicity in the upper crust. According to this
view, major earthquakes would reflect this tectonic driving, while
triggered earthquakes would be mostly ``witnesses'' (and relatively
minor ``actors'') of the physics of stress redistribution and
delayed rupture nucleation (according to stress corrosion, damage theory
and/or state and velocity dependent friction). 

We think that reality is probably a mixture of critical damage build-up
and stress accumulation due to the lower crust loading.
While it seems unrealistic to neglect (as often done) the strength
of the plastic and plastic-ductile lower crust and its mechanical
coupling with the upper brittle crust, it also seems that 
rupture in highly heterogeneous rocks does occur generically 
according to the critical rupture scenario. The important point
for our purpose is that both mechanisms combine to produce
large scale coherent stress loading and damage processes.

\section{Putting together the electric battery and critical earthquake concepts
for large scale electro-magnetic signatures}

When stress is applied, dislocations that move through the mineral grains
cause the PHPs to break up and to release p-hole charge carriers. The
p-holes ``live'' in the valence band. They travel by passing from oxygen
anion to oxygen anion using energy levels provided at the upper edge of
the valence band. Following their concentration gradient between
stressed and unstressed rock, the p-holes spread out of the stressed rock
volume. They propagate as charge waves with an estimated group velocity
on the order of $100-300$ m/s (Freund, 2002). At the
laboratory scale, this phenomenon is observed as 
the arrival of a charge cloud building up a concentration of
positive charges at the surface of the sample, with electric fields of
the order of $10^5 - 10^6$ V/cm. Such high fields may lead to
field-ionization of air molecules, the injection of ions into the air
and eventually to corona discharges accompanied by light emissions that have
indeed been measured (Freund, 2000).

Therefore, if there is a large scale region subjected to increasing
damage at the microscopic scale, it will act as a net source of p-holes
charge carriers flowing out to the surface and creating a net positive
group potential. This charge accumulation can be counteracted 
in two ways: (1) by an outflow of electrons from the stressed rock volume and (2) by an
inflow of protons possibly available from free water present in the
crust (Freund et al., 2004). In a steady state, the two charge fluxes will balance and no net
effect should be observed. Also, when no percolating conduction path
exists, the charge accumulation increases as in the charge of a battery
with no possibility for conduction.
When a large earthquake becomes incipient,
the critical model predicts that the damage should accelerate at the
microscopic scale (which may not be usually 
detectable by increasing seismicity at observable levels
or with other mechanical signatures), 
corresponding to an increasingly stronger source of
p-holes charge carriers
released from the rupture of peroxy bonds. In these circumstances,
one might be able to observe the effects associated with 
a sudden increase of outflow currents, when the conditions are such that
the circuit loop closes on itself allowing p-holes and electrons in the
battery volume to flow out. These p-hole and 
electron currents are coupled through their respective electric fields 
and, hence, fluctuating. Another effect of the stress activation
of these electronic charge carriers is the build-up of a net positive
ground potential over the broad regional areas surrounding an impending
large earthquake. 

Essentially all the electro-magnetic manifestations reported
anecdotically in the literature (Park et al., 1993) can be rationalized
by our theory, once we understand that igneous and high-grade metamorphic rocks
contain dormant electronic charge carriers. Stress ``awakens'' these
charge carriers, p-holes and electrons. As a result the stressed rock
volume turns into a battery from where electric currents can flow out. 
For these currents to flow in any sustained manner, however, certain
conditions have to be fulfilled. If the build-up of stress occurs
somewhere in the cool, brittle upper to middle crust, the p-holes can
always spread out laterally into the surrounding unstressed rocks. By
contrast, the electrons are blocked from entering the surrounding
unstressed rocks. They require a conductivity path downward into the
hot, n-type conductive lower crust. Once this downward connection is
established, the circuit loop closes allowing both currents, electrons
and p-holes, to flow out. Depending on conditions that are still poorly
characterized, the outflows may occur in bursts giving rise to pulses of
low-frequency electromagnetic radiation. In addition, the two outflow
currents can be expected to couple via the electric fields that the
propagating charges build up. Such a coupling should give rise to
fluctuations, which would be another source of low-frequency
electromagnetic emissions.

Small earthquakes at depth within the crust are expected to generate
individually electric signals which are too weak to be detected. 
But their collective electric effect
during the build-up phase towards the critical point, in which
many small events below the detection threshold are triggered, may be significant
and detectable at the surface. Indeed, there are reasons to believe
(Sornette and Werner, 2005a,b)
that there is an immense swarm of small earthquakes below detection thresholds
which, notwithstanding their small size, may dominate the 
triggering of detectable earthquakes. 
Collectively, these events can produce the conditions for the buid-up
of the electric charges and their circulation. In this respect,
mine rockburst offer interesting analogs and an intermediate scaling range
to test this idea. First, the power law acceleration of Benioff strain
often reported for earthquakes (Keilis-Borok and Malinovskaya, 1964; Sykes and Jaum\'e, 1990;
Bufe and Varnes, 1993; Saleur et al., 1996; Bowman et al., 1998;
Jaum\'e and Sykes, 1999) is also a characteristic of the space-time
organization of rockbursts prior to their largest counterparts 
(Ouillon and Sornette, 2000). Second, it
was suggested that electro-magnetic radiation 
also exhibits a similar acceleration and can thus be useful for the
prediction of rockbursts (Frid and Vozoff, 2005). We also note
that the organization of seismicity prior to mainshocks seems
similar, with the same statistical signatures of accelerated Benioff strain
and Gutenberg-Richter distribution (Rabinovitch et al., 2001),
both for mine rockbursts as mentioned above, to earthquakes and all the way to the scale
of neutron stars, whose quakes are observed in the form of soft gamma-ray repeaters
(Kossobokov et al., 2001; Sornette and Helmstetter, 2002).

Our theory also opens the door toward understanding, at least in
principle, other pre-earthquake phenomena that have been widely reported
in the literature. Prior to major earthquakes, the ionosphere displays
remarkable perturbations (Molchanov et al. 1993; Molchanov and Hayakawa
1998b, c; Alperovich and Fedorov 1999; Chen et al. 1999; Cliverd et al.
1999; Liu et al. 2000, 2001; Pulinets et al. 2000; Sorokin et al. 2001;
Pulinets and Boyarchuk 2004). These perturbations extend over large
areas, on the order of 500-1000 km. They require changes of the electric
field at ground level, strong enough to induce a recognizable reaction
in the ionosphere.  If the p-hole charge carriers are able to spread out
of a stressed rocks volume in the upper to middle crust, they can be
expected to reach the Earth's surface. If they get trapped at the
surface as laboratory experiments suggest (Freund, et al., 1993; Freund, et al., 1994), 
the surface will acquire a
positive charge. In the field, this would be equivalent to a change in
the Earth's ground potential over the region where the stress
accumulation takes place. The Earth-ionosphere system can be considered
a capacitor of which one plate, fixed at the Earth's surface, becomes
positively charged. The ionosphere, representing the other, flexible
plate, will then react and produce a mirror image of the charge
accumulation on the ground.

Other effects derive from the electric fields that are built up on a
microscopic scale at the rock-air interface when p-holes arrive at the
surface.  Theory predicts that the p-holes form charge layers at the
interface, $10-100$ nm thick, with charge densities sufficient to produce
surface potentials on the order of $0.1-1$ V (King and Freund, 1984). Experimentally, surface
potentials in the range of $50$ mV to $10-20$ V have been observed 
(Freund, et al., 2004). The
associated electric field, calculated for a flat surface (King and Freund, 1984), then fall
into the range of $5 \cdot 10^5-10^7$ V/cm. At sharp points, the electric fields
can be expected to reach values high enough to cause field ionization of
air molecules at the interface and injection of positive ions into the
atmosphere. These processes would in turn be consistent with dielectric
breakdown of the air and with glow or corona discharges, i.e. with
luminous phenomena and high-frequency electromagnetic emissions in the
range of tens of kHz.

Yet another widely reported pre-earthquake phenomenon may find its
physical explanation through a better understanding of the p-hole charge
carriers: the observation that, prior to major earthquakes in semi-arid
parts of the world, large areas of the land surface around the future
epicenter tend to emit infrared radiation equivalent to a temperature
increase on the order of $2-4^{\circ}$C (Dey and Singh, 2003; Srivastav, et al., 1997; 
Tronin, 1996; Tronin, et al., 2004; Qiang, et al., 1999;
Freund, et al., 2003; Ouzounov and Freund, 2004; Liu, et al., 2000;
Naaman, et al., 2001; Ohta, et al., 2001; Ondoh, 1998; Vershinin, et al., 1999;
Pulinets, et al., 2005; Trigunait et al., 2004; Yen, et al., 2004]. 
These ``thermal anomalies'' come and go
rapidly. After the earthquake and any aftershocks they disappear.
If p-holes spread out from a stressed rock volume, travel though thick
layers of rock and reach the Earth's surface, they can be expected to
recombine to form peroxy links similar to those that were broken deep
below during application of stress. During hole-hole recombination,
energy is released. A consequence of this process would be the formation
of vibrationally excited O-O bonds. These excited bonds can dissipate
their excess energy by emitting IR photons at the characteristic
frequencies of the O-O stretching vibration or by transferring energy
onto their Si-O and Al-O neighbors causing them to become excited and to
emit at their characteristic frequencies. Laboratory experiments have
produced results that are consistent with this line of reasoning. They
show that, when one end of a large block of rock is subjected to stress,
a free surface $50$ cm away instantly begins to emit in the infrared. The
emission spectrum is composed of sharp bands at the characteristic
frequencies of O-O, Si-O and Al-O vibrations (Freund et al., 2003).

A further piece of evidence for an outflow of p-holes from a stressed
rock volume is provided by the serendipitous observation of strong
magnetic field variations along the $110$ km long Chelungpu Fault in
Taiwan that ruptured during the M=7.7 Chi-Chi earthquake on Sept. 21,
1999. For seven week before the main event and for several weeks during
the intense aftershock period, the local magnetometer network recorded
rapid, pulse-like variations of the magnetic field, each lasting for
several hours (Yen et al., 2004). Telluric currents on the order to
$10^6$ Amp would have been necessary to produce such magnetic field
excursions (Freund et al., 2004).

Another consequence of our theory is the following. As the crust
``battery'' produces a current, the outflow of p-holes through the rock
is equivalent to electrons hopping back into the stressed rock volume.
These electrons will reconstitute the positive hole pairs inside the
stressed rock volume.  Therefore, the battery is (theoretically)
inexhaustible: the current can run until the battery is ``empty.'' As
stress is varying in time due to intermittent earthquakes and the slow
tectonic loading, the rock again charges up. The battery should be
reusable many, many times, paralleling the many, many seismic cycles
over millions of years of a given tectonic plate. In this vein, let us
mention a recent study supporting the idea that electric signals in the
crust (ultra-low frequency ground electric signals from stations
operated by the China Seismological Bureau over the last 20 years) have
a statistically significant pre-seismic component and are thus linked
with the preparatory stage before earthquakes (Zhuang et al., 2005).

An order-of-magnitude estimation, extrapolating from the laboratory
at the scale of rock blocks of 1 meter long, suggests very large currents
in the range of thousands to millions of amperes flowing out of the
hundreds to thousands of cubic kilometers of rock, which experience
increasing levels of stress prior to the catastrophic rupture. 
Conceptually, these strong currents can be concentrated
in narrow conducting paths. This begs the question
of why we do not observe these currents and EM radiations more easily.
To address this question, recall that our proposed concept is that of 
the crust acting like a battery being electrically 
``charged'' by the application of stress. Inside the
stressed rock volume, both electrons and defect electrons
(p-holes) are activated.
The p-holes can spread out into the surrounding unstressed rocks because
p-holes move along the upper edge of the valence band (which is
dominated by energy states of O2s and O2p character). In other words,
unstressed rock are (every so slightly) p-type conducting. We believe
that, as a result, p-holes can travel very far.
The electrons, however, are stuck inside the stressed rock volume. They
cannot leave it unless an n-type contact is provided, which is easily
done in the laboratory by sticking a Cu tape on the stressed rock volume or
using the steel pistons as contacts.  In the laboratory, a wire from the
stressed rock volume is run to any place on the unstressed rock, which closes
the loop and allows the battery to discharge.  Out in nature, we propose that the
n-type contact can be provided by rocks deeper in the
crust that exceed $\sim 500^{\circ}$C and thereby become n-type conductive.
Another possibility is through the percolation of fractures filled with small
amounts of conductive brine, which is expected to occur close to the critical point.
Therefore, the answer to the question, ``why we don't ``see'' powerful currents
more often and more easily'', is that the battery is there, ready to release
hundreds of thousands of amperes per km$^3$, but the electric loop is normally not
closed.  The loop closes only when massive stresses build up that reach
from the upper crust down to the hot lower crust or when damage percolates
through a conductive path, closing the electric circuit. 

Our theory is falsifiable (Kagan, 1999) as it makes specific tests that
can be refuted. Our preferred crucial test would consist in deploying a
large array of stations capable of measuring the potential difference of
the crust at the scale of tens of kilometers or more. To the best of our
knowledge, such measurements have never been done before but constitute
one of the best probe for detecting anomalous accumulation of positive
charges at locations related to large inpending earthquakes.

{\bf Acknowledgements}: We thank Xinglin Lei and another
anonymous referee for useful
comments which helped improve the manuscript.

{} 

\clearpage

\begin{figure}
\caption{\label{fig1} Simple picture sequence for the conversion of
hydroxyl pairs in MgO to peroxy (PHP) plus $H_2$ and the ``tearing apart'' of
the PHP to generate a p-hole.  The circles indicate the delocalization
of the charge associated with the p-hole and with the defect that is left
behind at the site of the new broken peroxy bond. The case of Si-OO-Si in
silicates, while more complex to represent, proceeds analogously.
} 
\end{figure}

\clearpage

\begin{figure}
\caption{\label{fig2} Brittle and ductile response at the level of small
rock volumes. (a) Increasing stress in the brittle regime causes
dislocation movement, coalescence of dislocations to microfractures and
merging of microfractures to cracks. (b) Increasing stress in the
ductile regime causes plastic deformation but, if the stresses become
very high and increase rapidly, even ductile materials can develop cracks.} 
\end{figure}

\clearpage

\begin{figure}
\caption{\label{fig3} Cross section through two crustal blocks
colliding at a constant speed. (a) Strain versus time. (b) Stress versus
time for a brittle material being deformed at a constant speed. (c)
Viscosity decreasing with increasing depth and temperature. (d) Volume
of rock undergoing brittle fracture as a function of time assuming
deformation at a constant speed and, hence, increasing levels of stress.} 
\end{figure}


\begin{thebibliography}{}

\bibitem{Allegre} All\`egre, C.J., J.L. Le Mouel, and A. Provost, 1982.
Scaling rules in rock fracture and possible implications for earthquake
predictions, Nature, 297, 47-49.

\bibitem{Alperovich} Alperovich, L., and E. Fedorov, 1999. Perturbation of atmospheric
conductivity as a cause of the litosphere-ionopshere interaction, in
Atmospheric and Ionospheric Electromagnetic Phenomena Associated with
Earthquakes, edited by M. Hayakawa, pp. 591-596, Terra Sci. Publ.,
Tokyo, Japan.

\bibitem{ASrecent1} Andersen, J.V. and D. Sornette, 2005. Predicting Failure
using Conditioning on Damage History: Demonstration on Percolation and
Hierarchical Fiber Bundles, Phys. Rev. E 72, 056124.

\bibitem{Andersen} Andersen, J.V., D. Sornette and K.-T. Leung, 1997.
Tri-critical behavior in rupture induced by disorder, Phys. Rev. Lett.
78, 2140-2143.

\bibitem{Ani} Anifrani, J.-C.,  C. Le Floc'h, D. Sornette and B.
Souillard, 1995. Universal Log-periodic correction to renormalization group
scaling for rupture stress prediction from acoustic emissions, J.Phys.I
France 5, 631-638.

\bibitem{Ashby1} Ashby, M.F. and D.R.H. Jones, 1980. Engineering Materials 1:
An Introduction to their Properties and Applications, Pergamon Press.

\bibitem{BakTang} Bak, P. and C. Tang, 1989. Earthquakes as a self-organized
critical phenomenon, J. Geophys. Res., 94, 15635-14637.

\bibitem{Bell} Bell, D.R., and G.R. Rossman, 1992. Water in earth's mantle:
The role of nominally anhydrous minerals., Science 255, 1391-1397.

\bibitem{Bianchi} Bianchi, R. and al., e., 1984. Radiofrequency
emissions observed during macroscopic hypervelocity impact experiments.
Nature, 308, 830-832.

\bibitem{Boumezard} Bouchaud,  J.-P. and M. Mezard, 1994.
Self-induced quenched disorder: a model for the glass transition, 
J. Phys. I France 4, 1109-1114.

\bibitem{Levin3} Bowman, D.D. and G.C.P. King, 2001a. Stress transfer and
seismicity changes before large earthquakes, C. R. Acad. Sci. Paris,
Sciences de la Terre et des plan\`etes, 333, 591-599.

\bibitem{Levin4} Bowman, D.D. and G.C.P. King, 2001b. Accelerating seismicity
and stress accumulation before large earthquakes, Geophys. Res. Lett.,
28, 4039-4042.

\bibitem{Bowman} Bowman, D.D., G. Ouillon, C.G. Sammis, A. Sornette and
D. Sornette, 1998. An Observational test of the critical earthquake concept,
{\it J.Geophys. Res., 103,} 24359-24372.

\bibitem{intercrit2} Bowman, D.D. and C.G. Sammis, 2004. Intermittent
criticality and the Gutenberg-Richter distribution, Pure Appl. Geophys.,
104, 1945-1956.

\bibitem{Brehm1} Brehm, D.J. and L.W. Braile, 1999a. Refinement of the modified
time-to-failure method for intermediate-term earthquake prediction, J.
Seismology 3, 121-138.

\bibitem{Brehm2} Brehm, D.J. and L.W. Braile, 1999b. Intermediate-term
earthquake prediction using the modified time-to-failure method in
southern California, Bull. Seism. Soc. Am. 89, 275-293.

\bibitem{Bufe93} Bufe, C.G., and D.J. Varnes, 1993. Predictive modelling of
the seismic cycle of the greater San Francisco bay region, J. Geophys.
Res., 98, 9871-9883.

\bibitem{Bufeetala} Bufe, C.G., S.P. Nishenko, and D.J. Varnes, 1994.
Seismicity trends and potential for large earthquakes in the
Alaska-Aleutian region, PAGEOPH, 142, 83-99.

\bibitem{Chen} Chen, Y.I., J.Y. Chuo, J.Y. Liu, and S.A. Pulinets, 1999. 
A statistical study of ionospheric precursors of strong earthquakes in
the Taiwan area,  XXVI URSI General Assembly,
Toronto, 13-21 Aug. 1999, Abstracts, p.745.

\bibitem{Cliver}  Clilverd,ÊM.A., Rodger,ÊC.J. and Thomson,ÊN.R., 1999. Investigating
seismoionospheric effects on a long subionospheric path, J. Geophys.
Res., 104, 171-179.

\bibitem{Con1} Contoyiannis, Y.F., P.G. Kapiris and K.A. Eftaxias, 2005.
Monitoring of a preseismic phase from its electromagnetic precursors,
Phys. Rev. E 71, 066123.

\bibitem{debatevan} Debate on VAN (special issue), 1996. Geophys. Res. Lett. 23, 1291-1452.
 
\bibitem{Derr} Derr, J.S., 1973. Earthquake lights: a review of observations and 
present theories. Bull. Seismol.
Soc. Amer., 63, 2177-21287.

\bibitem{Dey} Dey, S., and R. P. Singh, 2003. Surface latent heat flux as 
an earthquake precursor, Natural Hazards and Earth System Sciences, 3, 749-755.

\bibitem{Dickinson} Dickinson, J.T., L.C. Jensen, M.R. McKay, and F.T.
Freund, 1986. The emission of atoms and molecules accompanying fracture of
single-crystal MgO, J. Vac. Sci. Technol. 4 1648-1652.

\bibitem{Dieterich} Dieterich, J.H., 1992. Earthquake nucleation on faults
with rate-dependent and state-dependent strength, Tectonophysics 211
115-134.

\bibitem{Edwards} Edwards, A.H. and Fowler, W.B., 1982. Theory of the
peroxy-radical defect in a$-SiO/$sub 2/, Phys. Rev. B 26, 6649-6660.

\bibitem{Eft0} Eftaxias, K., P. Kapiris, E. Dologlou, J. Kopanas, N. Bogris, 
G. Antonopoulos, A. Peratzakis and V. Hadjicontis, 2002. EM anomalies
before the Kozani earthquake: a study of their behavior through laboratory
experiments, Geophys. Res. Lett. 29 (8), 10.1029/2001GL013786.

\bibitem{Eft} Eftaxias, K., P. Frangos, P. Kapiris, J. Polygiannakis, 
J. Kopanas, A. Peratzakis, P. Skountzos,
and D. Jaggard, 2004. Review and a model of pre-seismic electromagnetic emissions 
in terms of fractal electrodynamics, Fractals 12, 243-273. 

\bibitem{Ell} Ellsworth, W.L., A.G. Lindh, W.H. Prescott, and D.J. Herd, 1981.
The 1906 San Francisco earthquake and the seismic cycle, in Earthquake
Prediction: An International Review, Maurice Ewing Ser.,Maurice Ewing
Series,4, edited by D.W. Simpson, and  P.G. Richards, pp. 126-140, AGU,
Washington, D.C.

\bibitem{Freundim} Freund, F.T., 2000. Time-resolved study of charge generation
and propagation in igneous rocks, J. Geophys. Res. 105, 11001-11019.

\bibitem{Freund02} Freund, F.T., 2002. Charge generation and propagation in rocks, J.
Geodynamics, 33, 545-572.

\bibitem{Freund03} Freund, F.T., 2003. On the electrical conductivity structure of the stable
continental crust, J. Geodyn. 35, 353-388.

\bibitem{Freundwend} Freund, F.T., and H. Wengeler, 1982. The infrared spectrum
of OH--compensated defect sites in C-doped MgO and CaO single crystals,
J. Phys. Chem. Solids 43, 129-145.

\bibitem{FreundOber} Freund, F.T., and G. Oberheuser, 1986. Water dissolved in
olivine: a single crystal infrared study, J. Geophys. Res. 91, 745-761.

\bibitem{Freundho} Freund, F.T., and R. Ho, 1996. Organic matter supplied to a
planet by tectonic and volcanic activity, in Circumstellar Habitable
Zones, L.R. Doyle, ed. p. 71-98, Travis House Publ., Menlo Park, CA.

\bibitem{Freundetal} Freund, F.T., M.M. Freund, and F. Batllo, 1993. Critical
review of electrical conductivity measurements and charge distribution
analysis of magnesium oxide, J. Geophys. Res. 98, 22209-22229.

\bibitem{Freundetal93} Freund,ÊF.T., Freund,ÊM.M. and Batllo,ÊF., 1993. 
Critical review of electrical conductivity
measurements and charge distribution analysis of magnesium oxide, J.
Geophys. Res., 98, 22209-22229.

\bibitem{Freundetal94} Freund, F.T., E.-J. Wang, F. Batllo, L. Desgranges, C. Desgranges, 
and M.M. Freund, 1994. Positive hole-type charge carriers in oxide
materials, in Grain Boundaries and Interfacial Phenomena in Electronic
Ceramics, edited by L. M. Levinson, pp. 263-278, Amer. Ceram. Soc.,
Cincinnati, OH.

\bibitem{Freundetal02} Freund, F.T., M. D. Cash, and T. Dickinson, 2002. 
Hydrogen in rocks:  An energy source for deep
microbial communities, Astrobiology, 2, 83-92.

\bibitem{Freundsss} Freund, F.T., et al., 2003. Stimulated IR emission
from the surface of rocks during deformation, Fall Meeting, Amer.
Geophys. Union, San Francisco, CA, abstract T51E-0200.

\bibitem{Freundetal04} Freund, F.T., Takeuchi, A., Lau, 
B.W.S., Post, R., Keefner, J., Mellon, J. and Al-Manaseer, A., 2004.
 Stress-induced changes in the electrical
conductivity of igneous rocks and the generation of ground currents,
Terrestrial, Atmospheric and Oceanic Sciences (TAO), 15, 437-468.

\bibitem{Freund06} Freund, F.T., M.A. Salgueiro da Silva, R.W.S. Lau, A.
Takeuchi, and H.H. Jones, 2006a. Electric currents along earthquake faults and
the magnetization of pseudotachylite veins, Submitted Jan. 23, 2006 to
Tectonophysics, Special Issue "Mechanical and Electromagnetic Phenomena
Accompanying Pre-seismic Deformation: from Laboratory to Geophysical
ScaleÓ Guest Editors: K. Eftaxias, V. Sgrigna, T. Tchelidze.

\bibitem{Freund06_2} Freund, F.T., A. Takeuchi and R.W.S. Lau, 2006b.
Electric Currents Streaming out of Stressed Igneous Rocks Ð A Step Towards
Understanding Pre-Earthquake Low Frequency EM Emissions,
in press in Journal of Physics \& Chemistry of the Earth,
Special Issue: ``Recent Progress in Seismo Electromagnetics''.

\bibitem{Frid} Frid V., Vozoff K., 2005. Electromagnetic radiation induced by mining
rock failure, Int. J. Coal Geology, 64, 57-65.

\bibitem{Friebele} Friebele, E.J., D.L. Griscom, M. Stapelbroek, and
R.A. Weeks, 1979. Fundamental defect centers in glass: The peroxy radical in
irradiated high-purity fused silica, Phys. Rev. Lett. 42, 1346-1349.

\bibitem{Frost} Frost, H.J. and M.F. Ashby, 1982. Deformation-Mechanism
Maps: The Plasticity and Creep of Metals and Ceramics, Pergamon Press.

\bibitem{Fuju} Fujinawa, Y., Matsumoto, T., Iitaka, H. and Takahashi, S., 2001. 
Characteristics of the
Earthquake Related ELF/VLF Band Electromagnetic Field Changes, American
Geophysical Union, Fall Meeting 2001. AGU, San Franscisco, CA, pp., S42A-0616.

\bibitem{fuji} Fujinawa, Y. and Takahashi, K., 1990. Emission of electromagnetic 
radiation preceding the Ito
seismic swarm of 1989. Nature, 347, 376 - 378.

\bibitem{Garci} Garcimartin, A.,  Guarino, A.,  Bellon, L. and
Ciliberto, S., 1997. Statistical properties of fracture precursors, Phys. Rev.
Lett. 79, 3202-3205.

\bibitem{geralfkd} Gershenzon, N. and Bambakidis, G., 2001. Modeling of 
seismo-electromagnetic phenomena.
Russian J. Earth Scie., 3, 247-275.

\bibitem{gokh} Gokhberg, M.B., Morgounov, V.A., Yoshino, T. and Tomizawa, I., 1982. Experimental
measurements of electromagnetic emissions possibly related to earthquakes in Japan. J.
Geophys. Res., 87, 7824-7828.

\bibitem{hajkalkq} Hayakawa, M., 1989. Satellite observation of low 
latitude VLF radio noise and their association
with thunderstorms. J. Geomagn. Geoelectr., 41, 573-595.

\bibitem{Heimpel} Heimpel, M., 1997. Critical behavior and the evolution of
fault strength during earthquake cycles, Nature, 388, 865-868.

\bibitem{HS03} Helmstetter, A. and D. Sornette, 2003.
Foreshocks explained by cascades of triggered seismicity, 
J. Geophys. Res. 108 (B10), 2457 10.1029/2003JB002409 01.

\bibitem{Hik} Hickman, S.H., M.D. Zoback, C.A. Barton, R. Benoit, J. Svitek and R.
Summers, 2000. Stress and permeability heterogeneity within the Dixie Valley
geothermal reservoir: recent results from well 82-5, Proceedings,
Twenty-fifth workshop on Geothermal Reservoir Engineering, Stanford
University, Jan. 24-26, SGP-TR-165.

\bibitem{Huang1} Huang, Y., H. Saleur, C. G. Sammis, D. Sornette, 1998.
Precursors, aftershocks, criticality and self-organized criticality,
Europhysics Letters 41, 43-48.

\bibitem{Huang2} Huang, Y., H. Saleur and D. Sornette, 2000. Re-examination of
log-periodicity observed in the foreshocks of the 1989 Loma Prieta
earthquake, J. Geophysical Research 105, B12, 28111-28123.

\bibitem{Ivanov} Ivanov, B.A., Okulewskij, B.A. and Basilwvskij, A.T., 1976. 
Impulse magnetic field due to
shock induced polarization in rocks as a possible cause of magnetic field anomalies on
the moon, related to craters. Pisma v Asronomichelskij J., 2, 257-260.

\bibitem{Jaumesy} Jaum\'e, S.C. and Sykes L.R., 1999. Evolving towards a
critical point: A review of accelerating seismic moment/energy release
prior to large and great earthquakes, Pure and Applied Geophysics 155,
279-305.

\bibitem{Johsor98can} Johansen, A. and D. Sornette, 1998. Evidence of discrete
scale invariance by canonical averaging, Int. J. Mod. Phys. C 9, 433-447.

\bibitem{Johsor00} Johansen, A. and D. Sornette, 2000. Critical ruptures, 
Eur. Phys. J. B 18, 163-181.

\bibitem{Johnson} Johnston, M. J. S., 1997. Review of electric and magnetic fields
accompanying seismic and volcanic activity, Surveys in
Geophysics, 18, 441Ð475.

\bibitem{Jones} Jones, L.M., 1994. Foreshocks, aftershocks, and earthquake
probabilities: Accounting for the Landers earthquake, Bull. Seismol.
Soc. Am., 84, 892-899.

\bibitem{Jonmolnar} Jones, L.M., and P. Molnar, 1979. Some characteristics of
foreshocks and their possible relationship to earthquake prediction and
premonitory slip on faults, J. Geophys. Res., 84, 3596-3608.

\bibitem{Kagan} Kagan, Y.Y., 1999. Is earthquake seismology a hard,
quantitative science? Pure and Applied Geophysics 155, 233-258.

\bibitem{Kapiris} Kapiris, P.G., G.T. Balasis, J.A. Kopanas, G.N.
Antonopoulos, A.S. Peratzakis and K.A. Eftaxias, 2004a. Scaling similarities of
multiple fracturing of solid materials, Nonlinear Processes in Geophysics
11(1), 137-151.

\bibitem{Kapiris2} Kapiris, P.G., K. Eftaxias, and T. Chelidze,  2004b.
Electromagnetic Signature of Prefracture Criticality in Heterogeneous Media,
Phys. Rev. Lett. 92, 065702.  

\bibitem{Kathrein} Kathrein, H., F.T. Freund, and J. Nagy, 1984. $O^-$ ions and
their relation to traces of $H_2O$ and $CO_2$ magnesium oxide: an EPR
study., J. Phys. Chem. Solids 45, 1155-1163.

\bibitem{Keilis90} Keilis-Borok, V., 1990. The lithosphere of the Earth as a
large nonlinear  system. Geophys. Monogr. Ser. 60, 81-84.

\bibitem{Keietal} Keilis-Borok, V.I., L. Knopoff, I.M. Rotwain, and C.R.
Allen, 1988. Intermediate-term prediction of occuerrence times of strong
earthquakes, Nature, 335, 690-694.

\bibitem{KBM} Keilis-Borok, V. I. and Malinovskaya, L. N., 1964. 
One Regularity in the Occurrence of Strong Earthquakes, J. Geophys. Res. B 69,
3019Ð3024.

\bibitem{kingaq} King, C.-Y., 1983. Electromagnetic emission before earthquakes. Nature, 301, 377.

\bibitem{Kingfreund} King, B.V., and F.T. Freund, 1984. Surface charges and
subsurface space charge distribution in magnesium oxide containing
dissolved traces of water., Phys. Rev. B 29, 5814-5824.

\bibitem{Levin2} King, G.C.P. and D. D. Bowman, 2003. The evolution of
regional seismicity between large earthquakes, J. Geophys. Res., 108(B2)
2096, doi: 10.1029/2001JB000783.

\bibitem{Knopoff} Knopoff, L., T. Levshina, V.I. Keilis-Borok, and C.
Mattoni, 1996. Increased long-range intermediate-magnitude earthquake activity
prior to strong earthquakes in California, {\it J. Geophys. Res., 101,}
5779-5796.

\bibitem{kopy} Kopytenko, Y.A., Matiashvili, T.G., Voronov, P.M., Kopytenko, 
E.A. and Molchanov, O.A.,
1993. Detection of ultralow frequency emissions connected with the Spitak earthquake
and its aftershock activity, based on geomagnetic pulsation data at Susheti and Vardzia
observatories. Phys. Earth Planet. Inter., 77, 85-95.

\bibitem{Kosso} Kossobokov V., Keilis-Borok V.I., Cheng B., 2001. Similarities of
multiple fracturing on a neutron star and on the Earth, Physical Review
E. 61, 4, 3529-3533.

\bibitem{lei1} Lei, X.-L., K. Kusunose, O. Nishizawa, A. Cho and T.
Satoh, 2000. on the spatio-temporal distribution of acoustic emissions in two
granitic rocks under compression: the role of pre-existing cracks,
Geophys. Res. Letts 27, 1997-2000.

\bibitem{lei3} Lei, X.-L., K. Kusunose, T. Satoh, and O. Nishizawa, 2003. The
hierarchical rupture process of a fault: an experimental study, Physics
of the Earth and Planetary Interiors 137, 213-228.

\bibitem{lei2} Lei, X.-L., K. Masuda, O. Nishizawa, L. Jouniaux, L. Liu,
W. Ma, T. Satoh and K. Kusunose, 2004. Detailed analysis of acoustic emission
activity during catastrophic fracture of faults in rocks, Journal of
Structural Geology 26, 247-258.

\bibitem{Levin1} Levin, S.Z., C.G. Sammis, and D. D. Bowman, 20006. An
observational test of the stress accumulation model based on seismicity
preceding the 1992 Landers, CA earthquake, Tectonophysics, 413, 39-52.

\bibitem{Lindh} Lindh, A.G., 1990. The seismic cycle pursued, Nature, 348,
580-581.

\bibitem{Liuetal00} Liu, J. Y., et al., 2000. Seismo-ionospheric signatures prior to M³6.0
Taiwan earthquakes, Geophys. Res. Lett., 27, 3113-3116.

\bibitem{Liuetal01} Liu, J. Y., et al., 2001. Variations of ionospheric total electron
content during the Chi-Chi earthquake, Geophys. Res. Lett., 28,
1383-1386.

\bibitem{maz} Ma, Q.-z., Jing-yuan, Y. and Gu, X.-z., 2003. The
electromagnetic anomalies observed at Chongming station and the Taiwan
strong earthquake (M=7.5, March 31, 2002). Earthquake (in Chinese), 23, 49-56.

\bibitem{Main_review} Main I., 1996. Statistical physics, seismogenesis, and
seismic hazard, Reviews of Geophysics 34 (4), 433-462.

\bibitem{martlel} Martelli, G. and Smith, P.N., 1989. Light, radiofrequency 
emission and ionization effects
associated with rock fracture. Geophys, J. Internatl., 98, 397-401.

\bibitem{Martens} Martens, R., H. Gentsch, and F.T. Freund, 1976. Hydrogen
release during the thermal decomposition of magnesium hydroxide to
magnesium oxide, J. Catalysis 44, 366-372.

\bibitem{Martin} Martin, R.J., and M. Wyss, 1975. Magnetism of rocks and
volumetric strain in uniaxial failure tests, Pure Appl. Geophys. 113,
51-61.

\bibitem{Mogi} Mogi K., 1969. Some features of recent seismic activity in and
near Japan {2}: activity before and after great earthquakes, Bull. Eq.
Res. Inst. Tokyo Univ. 47, 395-417.

\bibitem{Mol1} Molchanov, O. A., and M. Hayakawa, 1998a. Subionospheric VLF signal
perturbations possibly related to earthquakes, J. Geophys. Res., 103,
17,489-417,504.

\bibitem{molcha} Molchanov, O.A. and Hayakawa, M., 1998b. On the 
generation mechanism of ULF seimogenic
electromagnetic emissions. Phys. Earth Planet. Int., 105, 201-220.

\bibitem{Mol2} Molchanov, O. A., et al., 1993. Observation by the Intercosmos-24
satellite of ELF-VLF electromagnetic emissions assocated with
earthquakes, Ann. Geophysicae, 11, 431-440.

\bibitem{PlaceMora3} Mora, P. and D. Place, 2000. Microscopic simulation of
stress correlation evolution: implication for the Critical Point
Hypothesis for earthquakes, proceedings of The Second ACES workshop,
October 15-20, 2000, Tokyo and Hokone, Japan
(\url{http://www.tokyo.rist.or.jp/ACES_WS2/})

\bibitem{MouraLei} Moura A., X.-L. Lei and O. Nishisawa, 2005.  Prediction
scheme for the catastrophic failure of highly loaded brittle materials
or rocks, Journal of the Mechanics and Physics of Solids, 53, 2435-2455.

\bibitem{MouraYu} Moura A. and V.I. Yukalov, 2002. Self-similar extrapolation
for the law of acoustic emission before failure of heterogeneous
materials, International Journal of Fracture 118 (3), L63-L68.

\bibitem{MouraYu2} Moura A. and V.I. Yukalov, 2002. Self-similar extrapolation
for the law of acoustic emission before failure of heterogeneous
materials, International Journal of Fracture 115 (1), L3-8.

\bibitem{Naa} Naaman, S., et al., 2001. Comparison of simultaneous
variations of the ionospheric total electron content and geomagnetic
field associated with strong earthquakes, Natural Hazards and Earth
System Sciences, 1, 53-59.

\bibitem{NASDA} NASDA-CON-960003, 1997. Abstracts of Int. Workshop on Seismo-Electromagnetics '97.

\bibitem{Nechad1} Nechad, H., A. Helmstetter, R. El Guerjouma and D.
Sornette, 2005a. Creep Ruptures in Heterogeneous Materials, Phys. Rev. Lett.
94, 045501.

\bibitem{Nechad2} Nechad, H., A. Helmstetter, R. El Guerjouma and D.
Sornette, 2005b. Andrade and Critical Time-to-Failure Laws in Fiber-Matrix
Composites: Experiments and Model, Journal of Mechanics and Physics
ofÊSolids (JMPS) 53, 1099-1127.

\bibitem{nitsan} Nitsan, U., 1977. Electromagnetic emission accompanying 
fracture of quartz-bearing rocks.
Geophys. Res. Lett., 4: 333-336.

\bibitem{Ohta} Ohta, K., et al., 2001. ULF/ELF emissions observed in Japan, possibly
associated with the Chi-Chi earthquake in Taiwan, Natural Hazards and
Earth System Sciences, 1, 37-42.

\bibitem{Onhoh} Ondoh, T., 1998. Ionospheric disturbances associated with great
earthquake of Hokkaido southwest coast, Japan of July 12, 1993, Phys.
Earth  Planet. Inter, 105, 261-269.

\bibitem{Ouillon} Ouillon, G., C. Castaing and D. Sornette, 1996. Hierarchical
scaling of faulting, J. Geophys. Res. 101, B3, 5477-5487.

\bibitem{ouilsor} Ouillon, G. and D. Sornette, 2000. The critical earthquake
concept applied to mine rockbursts with time-to-failure analysis,
Geophysical Journal International 143, 1-22.

\bibitem{Ouz} Ouzounov, D., and F.T. Freund, 2004. Mid-infrared
emission prior to strong earthquakes analyzed by remote sensing data,
Adv. Space Res., 33, 268-273.

\bibitem{PlaceMora1} Place, D. and P. Mora, 1999. The lattice solid model to
simulate the physics of rocks and earthquakes: Incorporation of
friction, J. Computational Physics 150, 332-372.

\bibitem{PlaceMora2} Place, D., P. Mora, S. Abe and S. JaumŽ, 2000,
Lattice solid simulation of the physics of earthquakes: the model,
results and directions, in: The Physics of Earthquakes, eds. Rundle,
J.B., Turcotte, D.L. \& Klein, W. (Am. Geophys. Union, Washington), in
press (2000).

\bibitem{Park} Park, S.K., Johnston, M.J.S., Madden, T.A., Morgan, F.D.
et al., 1993. Electromagnetic precursors to earthquakes in the ULF band -
Review of observations and mechanisms, Reviews of Geophysics 31, 117-132.

\bibitem{Pulinetsboy} Pulinets, S. and K. Boyarchuk, 2004. 
Ionospheric Precursors of Earthquakes, 350 pp. (Springer, Heidelberg.)

\bibitem{pulinets} Pulinets, S.A., Legen'ka, A.D. and Alekseev, V.A., 1994. 
Pre-earthquakes effects and their
possible mechanisms, Dusty and Dirty Plasmas, Noise and Chaos in Space and in the
Laboratory. Plenum Publishing, New York, pp. 545-557.

\bibitem{Pulinets1} Pulinets, S., et al., 2005. Thermal, atmospheric and ionospheric
anomalies around the time of Colima M7.8 earthquake of January 21, 2003,
Annales Geophysicae, submitted.

\bibitem{Pulinets2} Pulinets, S. A., et al., 2000. Quasielectrostatic model of
atmosphere-thermosphere-ionosphere coupling, Adv. Space Res., 26,
1209-1218.

\bibitem{Qiang} Qiang, Z., et al., 1999. Satellite thermal infrared brightness
temperature anomaly image: short-term and impending earthquake
precursors, Science in China, Series D: Earth Sciences, 42, 313-324.

\bibitem{rabi}  Rabinovitch A., Frid V., Bahat D., 2001. Gutenberg-Richter-type
 relation for laboratory fracture-induced electromagnetic radiation.
 Physical Review E. 65, 011401.

\bibitem{Raleigh} Raleigh, C.B., K. Sieh, L.R. Sykes, and D.L. Anderson, 1982.
Forecasting southern California earthquakes, Science, 217, 1097-1104.

\bibitem{Regenauer-Lieb} Regenauer-Lieb, K. and D.A. Yuen,  2003. Modeling
shear zones in geological and planetary sciences: solid- and
fluid-thermalÐmechanical approaches, Earth-Science Reviews 63, 295-349.

\bibitem{Ricci} Ricci, D., et al., 2001. Modeling disorder in amorphous
silica with embedded clusters: The peroxy bridge defect center, Physical
Review B, 64, 224104-224108.

\bibitem{Rossman} Rossman, G.R., 1996. Studies of OH in nominally anhydrous
minerals, Phys. Chem. Minerals 23, 299-304.

\bibitem{Roux}  Roux, S., Hansen, A., Herrmann, H., Guyon, E., 1988. Rupture
of heterogeneous media in the limit of infinite disorder, J. Stat. Phys.
52, 237.

\bibitem{Sahimi} Sahimi, M. and S. Arbabi, 1996. Scaling laws for fracture of
heterogeneous materials and rock, Phys. Rev. Lett. 77, 3689-3692.

\bibitem{intercrit1} Sammis, C.G., D D. Bowman, and G.C.P. King, 2004.
Anomalous seismicity and accelerating moment release preceding the 2001
and 2002 earthquakes in northern Baja California, Mexico, Pure Appl.
Geophys., 161, 2369-2378.

\bibitem{Saleuretal} Saleur, H., C.G. Sammis, and D. Sornette, 1996. Discrete
scale invariance, complex fractal dimensions, and log-periodic
fluctuations in seismicity, J. Geophys. Res., 101, 17,661-17,677.

\bibitem{SS02_PNSA} Sammis, S.G. and D. Sornette, 2002. Positive Feedback,
Memory and the Predictability of Earthquakes, Proceedings of the
National Academy of Sciences USA, V99 SUPP1, 2501-2508.

\bibitem{Shamir} Shamir, G., M.D. Zoback, and F.H. Cornet, 1990. Fracture-induced stress
heterogeneity: Examples from the Cajon Pass scientific drillhole near
the San Andreas fault, California, In: Rock Joints (Proceedings Int'l.
Conf.on Rock Joints: Norwegian Geotechnical Institute, Leon, Norway,
June 7-8, 1990), 719-724, N. Barton \& O. Stephannson, eds., A.A.
Balkema, Rotterdam, 814 pp.

\bibitem{Smyth} Smyth, J.R., D.R. Bell, and G.R. Rossman, 1991. Incorporation
of hydroxyl in upper-mantle clinopyroxenes, Nature 351, 732-735.

\bibitem{SS89} Sornette, A. and D. Sornette, 1989, Self-organized
criticality and earthquakes, Europhys.Lett., 9, 197 (1989).

\bibitem{SorSor} Sornette, A. and D. Sornette, 1990. Earthquake rupture as a
critical point: Consequences for telluric precursors. Tectonophysics
179, 327-334.

\bibitem{Sorrevie} Sornette, D., 1999. Earthquakes: from chemical alteration
to mechanical rupture, Physics Reports 313, 238-292.

\bibitem{Sornettebook} Sornette, D., 2004. Critical Phenomena in Natural
Sciences (Chaos, Fractals, Self-organization and Disorder: Concepts and
Tools), 2nd edition, Springer Series in Synergetics, Heidelberg.

\bibitem{reviewYip} Sornette, D., 2005.
Statistical Physics of Rupture in Heterogeneous Media,
Article 4.4 in Volume I of the ``Handbook of Materials Modeling,''
ed. by S. Yip (Springer Science and Business Media), pp. 1313-1331
(\url{http://arxiv.org/abs/cond-mat/0409524})

\bibitem{Soranderjor} Sornette, D. and  J. V. Andersen, 1998. Scaling with
respect to disorder in time-to-failure, Eur. Phys. Journal B 1, 353-357.

\bibitem{ASrecent2}  Sornette, D. and J.V. Andersen, 2006. Optimal Prediction
of Time-to-Failure from Information Revealed by Damage, in press
in Europhys. Letts.
\url{http://arxiv.org/abs/cond-mat/0511134}

\bibitem{SH02} Sornette, D. and Helmstetter, A., 2002.
Occurrence of Finite-Time-Singularity in Epidemic Models of Rupture,
Earthquakes and Starquakes, Physical Review Letters 89 (15) 158501.

\bibitem{SorSam} Sornette, D. and C.G. Sammis, 1995. Complex critical
exponents from renormalization group theory of earthquakes :
Implications for earthquake predictions, J.Phys.I France 5, 607-619.

\bibitem{SW1} Sornette, D. and Werner, M., 2005a.
Constraints on the Size of the Smallest Triggering Earthquake from the
ETAS Model, Baath's Law, and Observed Aftershock Sequences,
J. Geophys. Res. 110, No. B8, B08304, doi:10.1029/2004JB003535.

\bibitem{SW2} Sornette, D. and Werner, M., 2005b.
Apparent Clustering and Apparent Background Earthquakes Biased by
Undetected Seismicity, J. Geophys. Res.,ÊVol.Ê110,ÊNo.ÊB9,ÊB09303,
10.1029/2005JB003621.

\bibitem{Sorokin} Sorokin, V. M., et al., 2001.  Electrodynamic model of the lower
atmosphere and the ionosphere coupling, Journal of Atmospheric and
Solar-Terrestrial Physics, 63, 1681-1691 (2001).

\bibitem{Srivastav} Srivastav, S. K., et al., 1997. Satellite data reveals pre-earthquake
thermal anomalies in Killari area, Maharashtra, Current Science, 72,
880-884 (1997).

\bibitem{stlaur} St-Laurent, F., 2000. The Saguenay, Qu\'ebec, earthquake 
lights of November 1988-January 1989.
Seismolog. Res. Lett., 71: 160-174.

\bibitem{Stauffer} Stauffer, D. and Aharony, A.,1992. Percolation theory
(Taylor and Francis, London).

\bibitem{Sykes90} Sykes, L. R. and Jaum\'e, S., 1990. Seismic activity on neighboring faults 
as a long-term precursor to large earthquakes in the San Francisco Bay area,
Nature (London) 348, 595Ð599.

\bibitem{Taylor} Taylor, P. and M. Purucker, 2002. Searching for a magnetic signature
from earthquakes in the ionosphere, 76 pp., Final Report,
GESS Requirements Definition Study.

\bibitem{Tocher} Tocher, D., 1959. Seismic history of the San Francisco
region, in San Francisco Earthquakes of 1957G.B. Oakeshott, pp. 39-48,
CDMG Special Report 57.

\bibitem{Trigunait} Trigunait, A.P., S. Pulinets and F. Li, 2004. Variations of the
ionospheric electron density during the Bhuj seismic event, Annales
Geophysicae, 22, 4123-4131.

\bibitem{Tronin1} Tronin, A. A., 1996. Satellite thermal survey - a new tool for the
studies of seismoactive regions, Internatl. J. Remote Sensing, 17,
1439-1455.

\bibitem{Tronin2} Tronin, A. A., et al., 2004. Thermal anomalies and well observations in
Kamchatka, International Journal of Remote Sensing, 25, 2649-2655.

\bibitem{tuskuda} Tsukuda, T., 1997. Sizes and some features of 
luminous sources associated with the 1995
Hyogo-ken Nanbu earthquake. J. Phys. Earth, 45: 73-82.

\bibitem{Varot} Varotsos, P., 2005. The physics of seismic electric signals (SES), TerraPub, Tokyo.
 
\bibitem{Vere} Vere-Jones, D, 1977. Statistical theories of crack propagation,
Mathematical Geology  9, 455-481.

\bibitem{cershi} Vershinin, E.F., Buzevich, A.V., Yumoto, K., Saita, K. and 
Tanaka, Y., 1999. Correlation of
seismic activity with electromagnetic emissions and variations in Kamchatka region. In:
M. Hayakawa (Editor), Atmospheric and Ionospheric Electromagnetic Phenomena
Associated with Earthquakes. Terra Sci. Publ., Tokyo, Japan, pp. 513-517.

\bibitem{Wilson} Wilson, K.G., 1979. Problems in physics with many scales of
length, Scientific American 241, August, 158-179.

\bibitem{yasui} Yasui, Y., 1973. A summary of studies on luminous 
phenomena accompanied with earthquakes.
Memoirs Kakioka Magnetic Observatory, 15, 127-138.

\bibitem{Yen} Yen, H.-Y. et al., 2004. Geomagnetic fluctuations during 
the 1999 Chi-Chi earthquake in
Taiwan. Earth Planets Space, 56, 39-45.

\bibitem{yashi1} Yoshida, S., Manjgaladze, P., Zilpimiani, D., Ohnaka, M. and Nakatani, M., 1994.
Electromagnetic emissions associated with frictional sliding of rock. In: M. Hayakawa
and Y. Fujinawa (Editors), Electromagnetic Phenomena Related to Earthquake
Prediction. Terra Scientific, Tokyo, pp. 307-322.

\bibitem{yoshocaa} Yoshino, T. and Tomizawa, I., 1989. Observation of 
low-frequency electromagnetic emissions as
precursors to the volcanic eruption at Mt. Mihara during November, Phys. Earth
and Planet.Interiors, 57, 32-39.

\bibitem{Yukamoura} Yukalov, V.I., A. Moura and H. Nechad, 2004. Self-similar
law of energy release before materials fracture, Journal of the
Mechanics \& Physics of Solids 52 (2), 453-465.

\bibitem{Zhourupr} Zhou, W.-X. and D. Sornette, 2002. Generalized q-Analysis
of Log-Periodicity: Applications to Critical Ruptures, Phys. Rev. E
046111, 6604 N4 PT2:U129-U136.

\bibitem{Zhuang} Zhuang, J., D. Vere-Jones, H. Guan, Y. Ogata and L. Ma, 2005.
Preliminary analysis of observations on the ultra-low frequency electric
field in a region around Beijing, Pure and Applied Geophysics 162,
1367-1396.

\bibitem{zlo} Zlotnicki, J. and Cornet, F.H., 1986. A numerical model of 
earthquake-induced piezomagnetic
anomalies. J. Geophys. Res., B91, 709-718.



\end{thebibliography}
\end{document}